\documentclass[]{aastex631}

\usepackage{amsmath}
\usepackage{mathtools} 

\begin{document}

\title[$H_0$ from DESI 2024]{Uncorrelated estimations of $H_0$ redshift evolution from DESI baryon acoustic oscillation observations}



\author[0009-0009-3583-552X]{X. D. Jia}
\affiliation{School of Astronomy and Space Science, Nanjing University, Nanjing 210093, China}

\author[0000-0002-5819-5002]{J. P. Hu}
\affiliation{School of Astronomy and Space Science, Nanjing University, Nanjing 210093, China}

\author[0000-0003-0672-5646]{S. X. Yi}
\affiliation{School of Physics and Physical Engineering, Qufu Normal University, Qufu 273165, China}

\author[0000-0003-4157-7714]{F. Y. Wang}\thanks{E-mail: fayinwang@nju.edu.cn}
\affiliation{School of Astronomy and Space Science, Nanjing University, Nanjing 210093, China}
\affiliation{Key Laboratory of Modern Astronomy and Astrophysics (Nanjing University), Ministry of Education, Nanjing 210093, China}



\begin{abstract}
The Dark Energy Spectroscopic Instrument (DESI) collaboration recently released the first year data of baryon acoustic oscillations (BAOs). Based on the five different tracers, the cosmological constraint shows a hint of deviation from the standard $\Lambda$CDM model. In this letter, we combine the DESI BAOs with other cosmic probes to constrain the evolution of Hubble constant as a function of redshift in the flat $\Lambda$CDM model. The non-parametric method is used to estimate the value of Hubble constant at different redshift bins. The correlation among different bins are removed by diagonalizing the covariance matrix. The joint data sample demonstrate a decreasing trend of Hubble constant with a significance of $6.4 \sigma$, which can naturally resolve the Hubble tension. To avoid statistical effects caused by the binning methods, we tested other three different binning methods and also found a decreasing trend. It may be due to dynamical dark energy or modified gravity.
\end{abstract}

\keywords{cosmological parameters; cosmology: theory}

\section{Introduction}
Exploring the expansion history of the Universe is crucial for enhancing our empirical understanding of cosmology. The discovery of accelerated expansion of the universe indicates that dark energy, with negative pressure, is the dominant component in the current universe \citep{1998AJ....116.1009R,1999ApJ...517..565P}. In recent years, the standard cosmological model $\Lambda$CDM has been widely supported by most cosmological observations \citep{PlanckCollaboration2020,2021PhRvD.103h3533A}. However, it is challenged by the Hubble tension \citep{Verde2019,Riess2020,Perivolaropoulos2022}. By combining the $\Lambda$CDM model with the cosmic microwave background (CMB) anisotropies measurements, the Planck Collaboration gives the Hubble constant $H_{0,z\sim1100} = 67.4 \pm 0.5$ km s$^{-1}$ Mpc$^{-1}$ \citep{PlanckCollaboration2020}. Using the local distance ladder, which relies on the Cepheid and type Ia supernovae (SNe Ia) measurements, the SH0ES team measured $H_{0,z\sim0} = 73.04\pm 1.04$ km s$^{-1}$ Mpc$^{-1}$ \citep{Riess2022}. The discrepancy between them is about $5 \sigma$ \citep{Riess2022}.

Recently, the Dark Energy Spectroscopic Instrument (DESI) collaboration has released its first round of cosmological constraints based on baryon acoustic oscillations (BAOs) \citep{2024arXiv240403002D}. The results provide hints of dynamic dark-energy behavior. For the $w_0w_a$CDM model, the dark energy equation of state is redshift dependent as $w(z) = w_0 +w_a\frac{z}{1+z}$. 
In this context, DESI reports a preference for $w_0 > -1$. When combined with Planck CMB data \citep{PlanckCollaboration2020} and SNe Ia, the preference for the $w_0w_a$CDM model over the $\Lambda$CDM model increases to $3.9 \sigma$.

The significant discrepancies with the $\Lambda$CDM model derived from DESI have triggered widespread discussion on dark energy. Several studies have sought to constrain cosmological parameters using the new DESI BAO data \citep{Bousis2024,2024arXiv240408633C,2024arXiv240504216C,2024arXiv240502168W,Pang2024,WangH2024}. The measurement in the luminous red galaxy seems to exhibit significant deviations from fiducial cosmology \citep{2024arXiv240502168W}. Some studies have opted to reanalyze cosmological constraints without considering them \citep{2024arXiv240408633C,2024arXiv240412068C,2024arXiv240413833W}. Even without considering the DESI BAO data, the current joint compilation of data still favors dynamical dark energy over the cosmological constant by $2\sigma$ \citep{Park2024}. By combining DESI, Planck and Pantheon+ data, the luminous red galaxy sample will have little impact on the constraint on ($w_0, w_a$) parameter estimation. The constraints from high-redshift and low-redshift data demonstrate hints of an evolution of Hubble constant \citep{Bousis2024}.

The marginal evidence on the research of the evolution of the Hubble constant is a possible way to solve the Hubble tension \citep{Kazantzidis2020,Hu2022,universe9020094,2023A&A...674A..45J,2024ApJ...975L..36H,2024A&A...681A..88H}. In the flat $\Lambda$CDM model, a descending trend in the Hubble constant has been found \citep{2020PhRvD.102j3525K,Wong2020,2021ApJ...912..150D,2022Galax..10...24D,2022PhRvD.106d1301O,OColgain2024,Malekjani2024}.Here, we use the novel non-parametric method proposed by \citet{2023A&A...674A..45J} to estimate the redshift evolution of the Hubble constant from the new DESI data. This non-parametric method is similar to the one used to constrain the dark energy equation of state \citep{2005PhRvD..71b3506H,Riess2007,Wang2011}.

This letter is organized as follows. The data is discussed in section \ref{Data}. The method and results are shown in section \ref{Methods}. Conclusions and discussion are given in section \ref{Conclusion and Discussion}.

\section{Data}\label{Data}
Recently, the first round of cosmological constraints based on BAO from the DESI collaboration has been released \citep{2024arXiv240403002D}. The constraints alone are consistent with the standard flat $\Lambda$CDM model. However, the combination of DESI, CMB and SNe Ia indicates a $2.6 \sigma$ discrepancy with the $\Lambda$CDM model. The BAO data from DESI 2024 provide a tight constraints on the cosmological models. In order to better study the evolution of the Hubble constant, we incorporate them into our original data sample presented in \cite{2023A&A...674A..45J}. Such a joint sample includes the latest observational results from multiple probes. Considering the redshift distribution of the sample and the binning method mentioned in Section \ref{Methods}, data with redshift higher than 1.5 are discarded and the remaining data are compiled in our research.

The first sample is the Hubble parameter sample, which contains 35 $H(z)$ measurements spanning a redshift range from $z = 0.07$ to $z = 1.965$. It has been used in previous literature multiple times \citep{2018ApJ...856....3Y,Moresco2020,2022MNRAS.513.5686C,2023A&A...674A..45J,Moresco2023}. The details of the sample are presented in \cite{Moresco2023}. They are derived using the cosmic chronometic technique, which is independent of cosmological models. The Hubble parameter is inferred by comparing the differential age evolution of galaxies at different redshifts with the formula $H(z) = -\frac{1}{1+z} \frac{d z}{d t}$ \citep{2002ApJ...573...37J}. The value of $d t$ is measured from the age difference between two passively evolving galaxies and $d z$ is the redshift interval between them. There is a covariance matrix between the data obtained using the method of analyzing specific spectral features \citep{Moresco2020}. 

Secondly, the old BAO data sample contains 12 measurements, which span the redshift range $0.122 \leq z \leq 2.334 $. The whole sample and their covariance matrix are shown in \cite{2023A&A...674A..45J}. The new sample is the DESI first year data release \citep{2024arXiv240403002D}. They are derived from these tracers: the bright galaxy sample, the luminous red galaxy sample, the emission line galaxy sample, the quasar sample and the Lyman-$\alpha$ forest sample. The compilation of compressed distance quantities $D_{\mathrm{M}}/r_{\mathrm{d}}$, $D_{\mathrm{H}}/r_{\mathrm{d}}$ and $D_{\mathrm{V}}/r_{\mathrm{d}}$ is used in this paper, as given in \cite{2024arXiv240403002D}. The BAO sample is calibrated by the CMB sound horizon scale distance $r_\mathrm{d} = 147.1$ Mpc at the end of the baryonic drag epoch \citep{PlanckCollaboration2020}. Following the method mentioned in \cite{2024arXiv240403002D}, we constructed a combined sample. The effective volume covered by the survey plays a crucial role in the sample selection process. The old BAO data have larger effective volume at $z < 0.6$, so we use the old data in place of the DESI points. DESI has larger effective volume at $z > 0.6$, which lead us to use the DESI results.

The data for the final sample is extracted from the Pantheon+ SNe Ia sample \citep{2022ApJ...938..113S}. The Pantheon+ sample consists of 1,701 light curves of 1,550 distinct SNe Ia spanning $0.01 \leq z \leq 2.26$. The uniform intrinsic luminosity makes it a standard candle, which is crucial for measuring the Hubble constant especially at low redshifts. Considering the dependence of low-redshift data points on peculiar velocity corrections, we only use SNe Ia with $z > 0.01$ here. After removing the data that do not meet the criteria, the sample used in this research includes 1,583 data points.

\section{Method and results}\label{Methods}
The value of Hubble constant is determined by extrapolating the Hubble parameter $H(z)$ from the observational data at higher $z$ to the local $z=0$, using a particular cosmological model. The redshift evolution of the Hubble constant can be studied by a non-parametric method \citep{2023A&A...674A..45J}, similar to the treatment of the equation of state of dark energy \citep{2005PhRvD..71b3506H,Riess2007,Jia2022}. To avoid imposing priors on the nature of the Hubble constant, we refrain from assuming that it follows specific functions. The value of Hubble constant are just allowed to remain a constant in each redshift bin.

\subsection{Methods}
Under the assumption of a piece-wise function, $H_0(z)$ can be represented as:
\begin{equation}\label{H0function}
    H_{0}(z)=\left\{\begin{array}{ll}H_{0, z_{1}} & \text { if } 0 \leq z<z_{1}, \\ H_{0, z_{2}} & \text { if } z_{1} \leq z<z_{2}, \\ \cdots & \cdots, \\ H_{0, z_{i}} & \text { if } z_{i-1} \leq z<z_{i}, \\ \cdots & \cdots, \\ H_{0, z_{N}} & \text { if } z_{N-1} \leq z<z_{N} .\end{array}\right.
\end{equation}
The parameter $N$ is the total number of redshift bins and $i$ represents the $i$-th redshift bin. The parameter $H_{0,z_i}$ represents the value of $H_0(z)$ in the $i$-th bin, spanning from $z_{i-1}$ to $z_i$. 

Basing on the flat $\Lambda$CDM model, the Hubble parameter is given by 
\begin{equation}\label{Hz}
H(z)=H_{0} \sqrt{\Omega_{m0}(1+z)^{3}+\Omega_{\Lambda0}} .
\end{equation}
Equation (\ref{Hz}) can be converted to an integral form
\begin{equation}\label{expansion integral}
H(z)= H_0\sqrt{\Omega_{m0}(1+z)^{3}+\Omega_{\Lambda0}} \\
=H_0\left(\int_{0}^{z}  \frac{3\Omega_{m0}(1+z^{\prime})^2}{2\sqrt{\Omega_{m0}(1+z^{\prime})^{3}+\Omega_{\Lambda0}}} dz^{\prime}+1\right) \\
= \int_{0}^{z}  \frac{H_03\Omega_{m0}(1+z^{\prime})^2}{2\sqrt{\Omega_{m0}(1+z^{\prime})^{3}+\Omega_{\Lambda0}}} dz^{\prime}+H_0.
\end{equation}
The evolution of the Hubble constant can be researched by combining Equation (\ref{H0function}) with Equation (\ref{expansion integral}). The final expression for the Hubble parameter is 

\begin{flalign}\label{Hz step}
    \begin{split}
        H\left(z_{i}\right) &= H_{0, z_{1}} \int_{0}^{z_{1}} \frac{3 \Omega_{m 0}(1+z)^{2}}{2 \sqrt{\Omega_{m 0}(1+z)^{3}+\Omega_{\Lambda 0}}}  \\
                    &+ H_{0, z_{2}} \int_{z_{1}}^{z_{2}} \frac{3 \Omega_{m 0}(1+z)^{2}}{2 \sqrt{\Omega_{m 0}(1+z)^{3}+\Omega_{\Lambda 0}}} \\ 
                    &+ \cdots \\ 
                    &+ H_{0, z_{i}} \int_{z_{i-1}}^{z_{i}} \frac{3 \Omega_{m 0}(1+z)^{2}}{2 \sqrt{\Omega_{m 0}(1+z)^{3}+\Omega_{\Lambda 0}}}+H_{0, z_{i}}.
    \end{split}
\end{flalign}
The last term $H_0$ in equation (\ref{expansion integral}) is replaced by $H_0(z)$. Therefore, it should be $H_0(z_i)$ when calculating $H(z_i)$. According to the definition in Equation (\ref{H0function}), it is equal to $H_{0,z_i}$ in this context. The value of $H_0(z)$ at low redshifts is determined by the evolution at high redshifts. The value of $H_0(z)$ will inform us whether it is evolving. It will revert to $H_0$ if there is no evolutionary trend.

The $\chi^2$ statistic method is used to estimate cosmological parameters with a set of parameters for $H_{0,z_i}$ as $\theta$ ($H_{0,z_i}$).
\begin{equation}\label{chi2}
\chi_{\theta}^{2}=\chi_{H(z)}^2+\chi_{BAO}^2+\chi_{SNe}^2.
\end{equation}

The value of $\chi_{H(z)}^2$ is 
\begin{equation}
\chi_{H(z)}^{2}=\sum_{i=1}^{N} \frac{\left[H_{\mathrm{obs}}\left(z_{i}\right)-H_{\mathrm{th}}\left(z_{i}\right)\right]^{2}}{\sigma^{2}_i} +\left[H_{\mathrm{obs}}\left(z_{i}\right)-H_{\mathrm{th}}\left(z_{i}\right)\right]\mathbf{C_{H(z)}}^{-1}\left[H_{\mathrm{obs}}\left(z_{i}\right)-H_{\mathrm{th}}\left(z_{i}\right)\right]^{T},
\end{equation}
where $H_{obs}(z_i)$ and $\sigma_i$ are the observed Hubble parameter and the corresponding 1$\sigma$ error. The covariance matrix $\mathbf{C_{H(z)}}$ can be found in \cite{Moresco2020}. It includes contributions from statistical errors, young component contamination, dependence on the chosen model, and stellar metallicity.

The value of $\chi^2_{BAO}$ is 
\begin{equation}
\chi_{BAO}^{2}= \left[\nu_{\mathrm{obs}}\left(z_{i}\right)-\nu_{\mathrm{th}}\left(z_{i}\right)\right]\mathbf{C_{BAO}}^{-1}\left[\nu_{\mathrm{obs}}\left(z_{i}\right)-\nu_{\mathrm{th}}\left(z_{i}\right)\right]^{T},
\end{equation}
where $\nu_{obs}$ is the vector of the BAO measurements at each redshift $z$ (i.e. $D_V\left(r_{s,{\rm fid}}/r_s\right), D_M/r_s, D_H/r_s, D_A/r_s$).

The value of $\chi^2_{SNe}$ is: 
\begin{equation}
\chi_{SNe}^{2}= \left[\mu_{\mathrm{obs}}\left(z_{i}\right)-\mu_{\mathrm{th}}\left(z_{i}\right)\right]\mathbf{C_{SNe}}^{-1}\left[\mu_{\mathrm{obs}}\left(z_{i}\right)-\mu_{\mathrm{th}}\left(z_{i}\right)\right]^{T}
.\end{equation}
The covariance matrix $\mathbf{C_{SNe}}$ contains the statistical matrix and the systematic covariance matrix. The parameter $\mu_{\mathrm{obs}}(z_i)$ is the distance module from the Pantheon+ sample.

The prior of $H_0$ is adopted as $H_0 \in$ [50, 80] $\textrm{km}~\textrm{s}^{-1} \textrm{Mpc}^{-1}$. A fiducial value of $\Omega_m = 0.315$ for the cosmic matter density is used during the fitting process. We analyze results constrained by each sub-sample individually: the sample from SNe Ia constrains $\Omega_m = 0.338 \pm 0.018$ \citep{Brout2022}, the sample from BAO constrains $\Omega_m = 0.297 \pm 0.012$ \citep{2024arXiv240403002D}, the sample from $H(z)$ measurements constrains $\Omega_m = 0.355^{+0.113}_{-0.088}$ \citep{Moresco2023}. The final value of $\Omega_m$ is determined to be 0.315, which is not only the average of the results from the SNe Ia and BAO samples, but also consistent with the result from CMB \citep{PlanckCollaboration2020}. The results from previous literature also demonstrate that such an approximation is reasonable for studying the evolutionary trend of the Hubble constant \citep{2023A&A...674A..45J}. The Markov Chain Monte Carlo (MCMC) code $emcee$ is used to derive the constraints \citep{Foreman-Mackey2013}.

\subsection{Results}\label{Results}
\begin{figure*}
    \centering
    \includegraphics[width=\textwidth,angle=0]{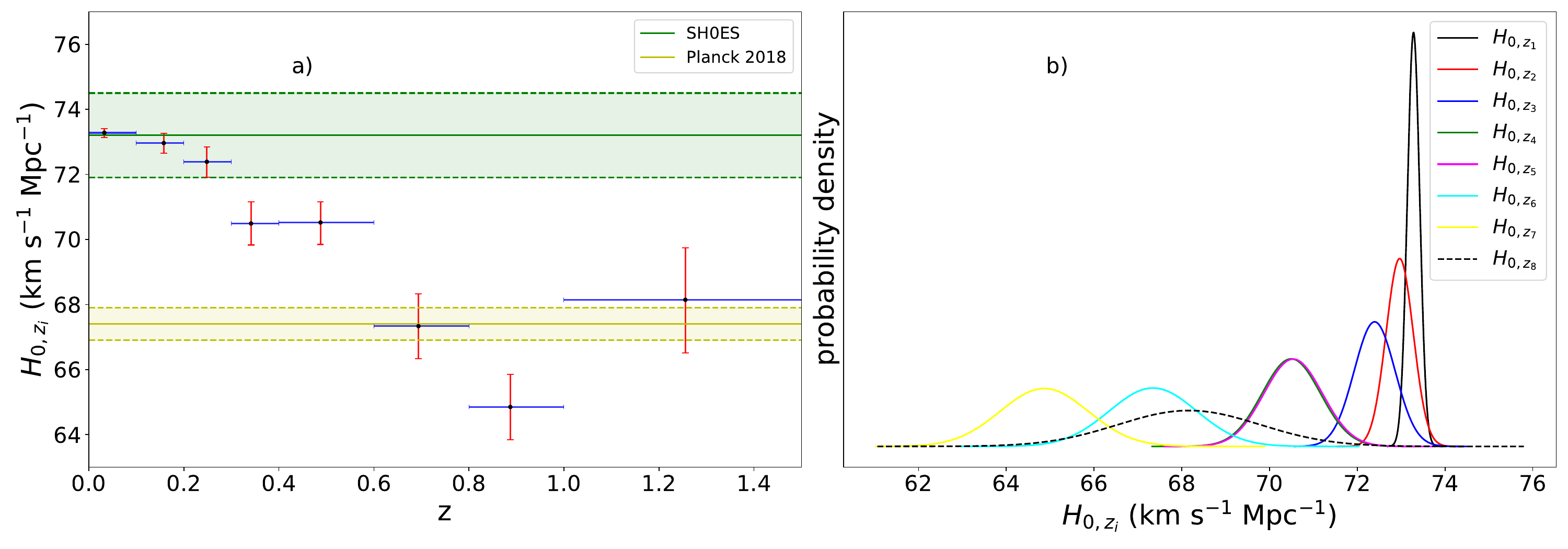}
    \caption{Fitting results of $H_{0,z}$ in eight redshift bins. Panel (a) shows the value of $H_{0,z}$ as a function of redshift. There is a clear decreasing trend with $6.4 \sigma$ significance between $z = 0$ to $z = 1.5$.  The green line gives $H_{0} = 73.04\pm 1.04$ km s$^{-1}$ Mpc$^{-1}$ from the local distance ladder and its 1$\sigma$ uncertainty \citep{Riess2022}. The yellow line is the value of $H_{0} = 67.4\pm 0.5$ km s$^{-1}$ Mpc$^{-1}$ from the CMB measurements and its 1$\sigma$ uncertainty \citep{PlanckCollaboration2020}. Panel (b) shows the probability density of $H_{0,z}$.}
    \label{Fbin9}
\end{figure*}

\begin{figure*}
        \includegraphics[width=1.0\textwidth,angle=0]{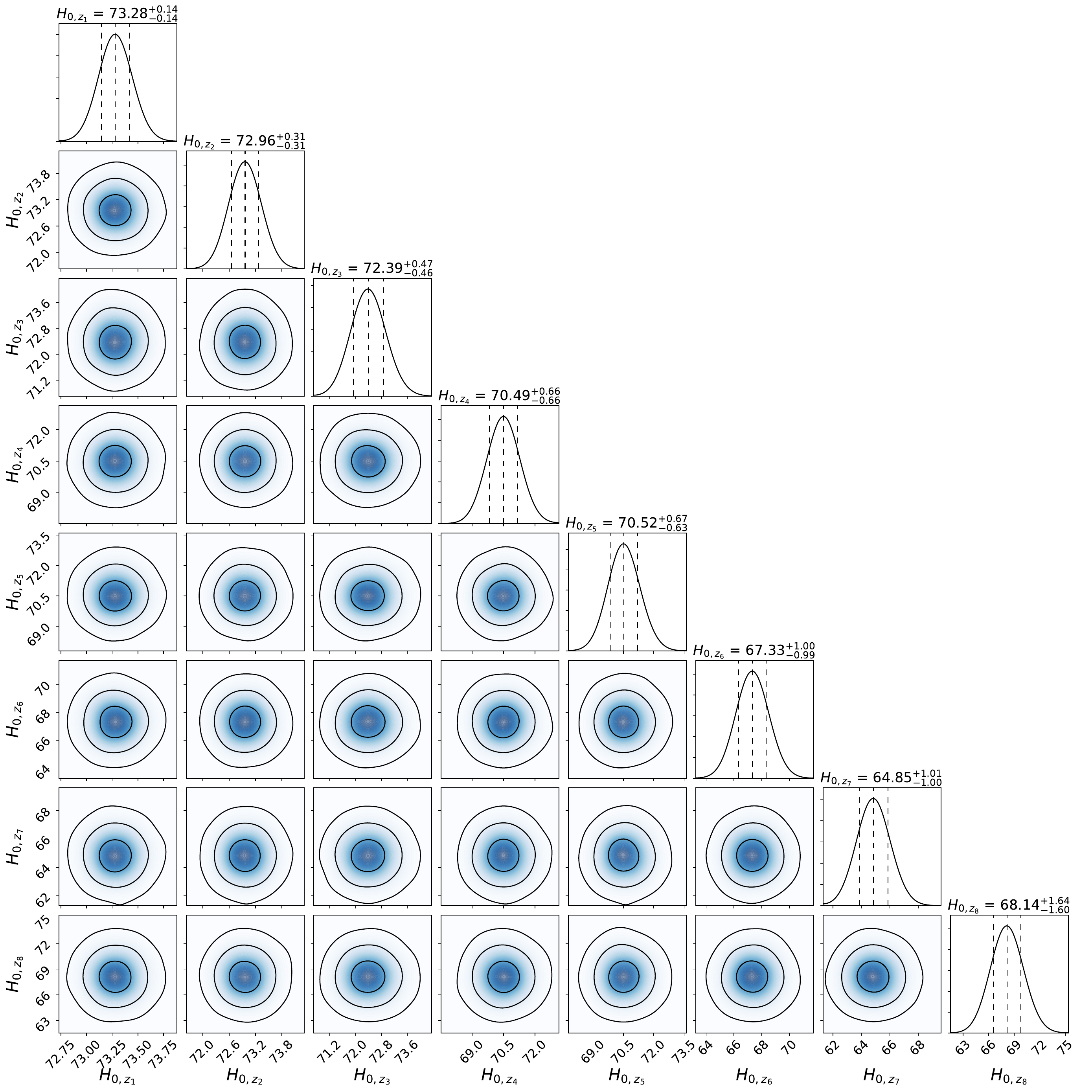}
        \caption{Corner plot of $H_{0,z}$ values in units of km s$^{-1}$ Mpc$^{-1}$. The panels on the diagonal show the 1D posterior probability distribution for each parameter obtained by marginalizing over the other parameters. The off-diagonal panels show two-dimensional projections of the posterior probability distributions for each pair of parameters, with contours to indicate 1$\sigma$ to 3$\sigma$ confidence levels.}
        \label{H9cor}       
\end{figure*}

In our previous work, two redshift-binning methods were adopted \citep{2023A&A...674A..45J}. Both methods revealed a decreasing trend. Considering that the number of high-redshift data points is much smaller than that of low-redshift data points, this will result in larger grouping intervals at high redshifts. In this study, we primarily focus on grouping by redshift. The number of samples in each redshift interval affects the error bars of the final results \citep{2001ApJ...559....9P,2013PhLB..726...72F,2017ApJ...835...26F}. Therefore, we try to keep the number of data in each redshift interval as similar as possible while maintaining equal redshift intervals. To study the redshift evolution of the Hubble constant, we divide the redshift range into eight intervals. The bins are equally spaced at low redshift. Due to the limitation in the number of samples, we are compelled to choose larger redshift intervals at higher redshifts. The upper boundaries of these eight bins are $z_i = 0.1, 0.2, 0.3, 0.4, 0.6, 0.8, 1.0$ and $1.5$. The evolution of the values of $H_{0,z}$ is given in Figure \ref{Fbin9} and Table \ref{T_H8}. The corner plots are shown in Figure \ref{H9cor}. The vertical error bars show the $1 \sigma$ uncertainty, while the horizontal error bars represent the range over which each $H_0(z)$ value spans. The fitting results at low redshifts are consistent with the value from the local distance ladder within the $1 \sigma$ confidence level \citep{Riess2022}. Although our results are higher than those from the Tip of the Red Giant Branch (TRGB) method, they still fall within the $1 \sigma$ range of agreement \citep{Freedman2021}. The value of $H_{0,z}$ at the last redshift bin is consistent with the value from CMB within the $1 \sigma$ confidence level \cite{PlanckCollaboration2020}. The decreasing trend in $H_0(z)$ is apparent from $z = 0$ to $z = 1.0$. We use the null hypothesis method to quantify the significance of the decreasing trend \citep{Wong2020,Millon2020,2023A&A...674A..45J}. We use the mean redshift of the data in each bin and the corresponding $H_{0,z}$ fitting results to form the point representing that bin. The significance is found to be $6.4 \sigma$. Comparing to our previous results \citep{2023A&A...674A..45J}, the new data sets from DESI provide tighter constraints on the evolution at high redshifts. The uncertainty of $H_{0,z}$ is reduced significantly.

\subsection{Test the decreasing trend}\label{Comparison}
The redshift evolution of the Hubble constant effectively alleviates the Hubble tension. In addition, we consider the impact of other potential factors. At first, we treat $\Omega_m$ as a free parameter in our fitting process. However, the posterior distribution of the parameters deviates from a Gaussian distribution starting from the second bin, due to the degeneracy between $\Omega_m$ and $H_{0,z}$. Such a non-Gaussian distribution will affect the calculation of the transformation matrix, which in turn impacts the final results. Considering the fact that the integration and summation over low redshift bins definitely affect the model fit in middle and higher redshift bins, these correlations need to be removed \citep{2023A&A...674A..45J}. The non-Gaussian posterior distribution affects the calculation of the transformation matrix, and results without removing the correlations cannot represent the true value of $H_{0,z}$. The quality of current data sample is insufficient to simultaneously constrain $\Omega_m$ and $H_{0,z}$. In order to test whether the value of $\Omega_m$ will affect the results, we examine the impact of different values of $\Omega_m$ on the $H_{0,z}$ results. The results are shown in Table \ref{T_Om}. For different values of $\Omega_m$, $H_{0,z}$ consistently exhibits a decreasing evolutionary trend. Due to the degeneracy between $\Omega_m$ and $H_{0,z}$, the value of $\Omega_m$ affects the result of $H_{0,z}$. Notably, the evolutionary trend remains consistent, which further indicates that the evolution of $H_{0,z}$ is driven by the model itself rather than by $\Omega_m$.

\begin{figure*}
    \centering
    \includegraphics[width=\textwidth,angle=0]{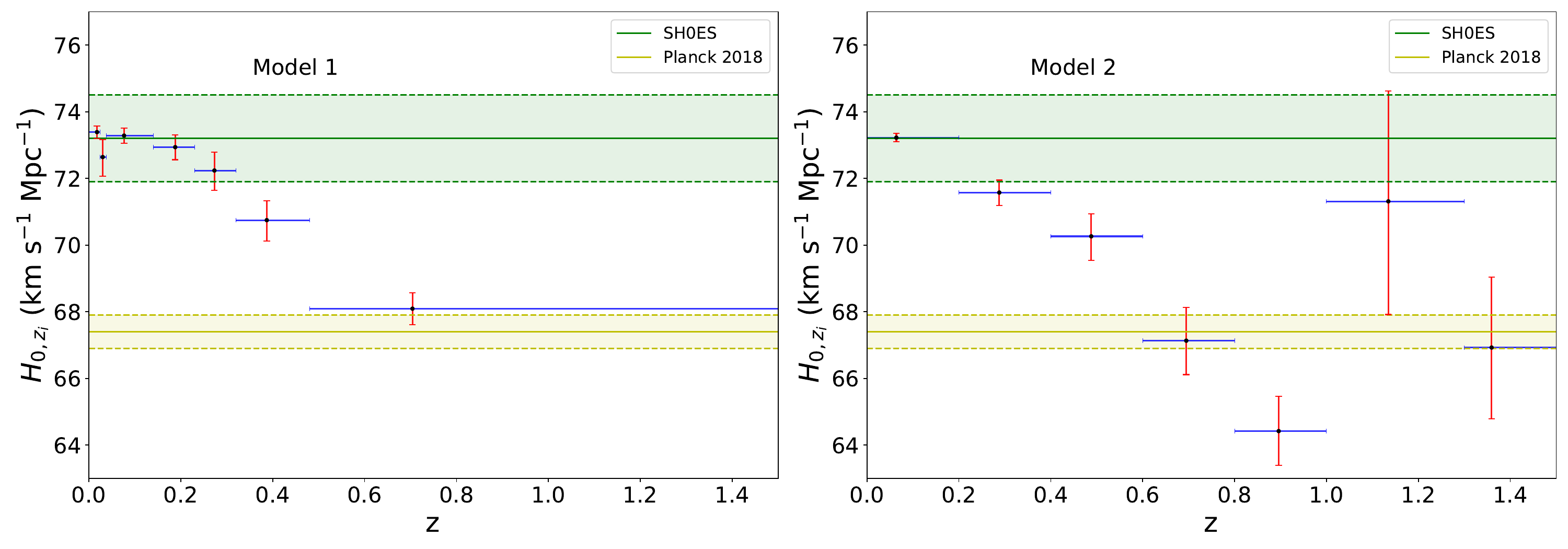}
    \caption{Fitting results of $H_{0,z}$ for Model 1 and Model 2. Panel (a) shows the value of $H_{0,z}$ as a function of redshift for Model 1, while panel (b) presents the value for Model 2. The green line gives $H_{0} = 73.04\pm 1.04$ km s$^{-1}$ Mpc$^{-1}$ from the local distance ladder and its 1$\sigma$ uncertainty \citep{Riess2022}. The yellow line is the value of $H_{0} = 67.4\pm 0.5$ km s$^{-1}$ Mpc$^{-1}$ from the CMB measurements and its 1$\sigma$ uncertainty \citep{PlanckCollaboration2020}.}
    \label{Fbin7}
\end{figure*}

\begin{figure*}
    \centering
    \includegraphics[width=\textwidth,angle=0]{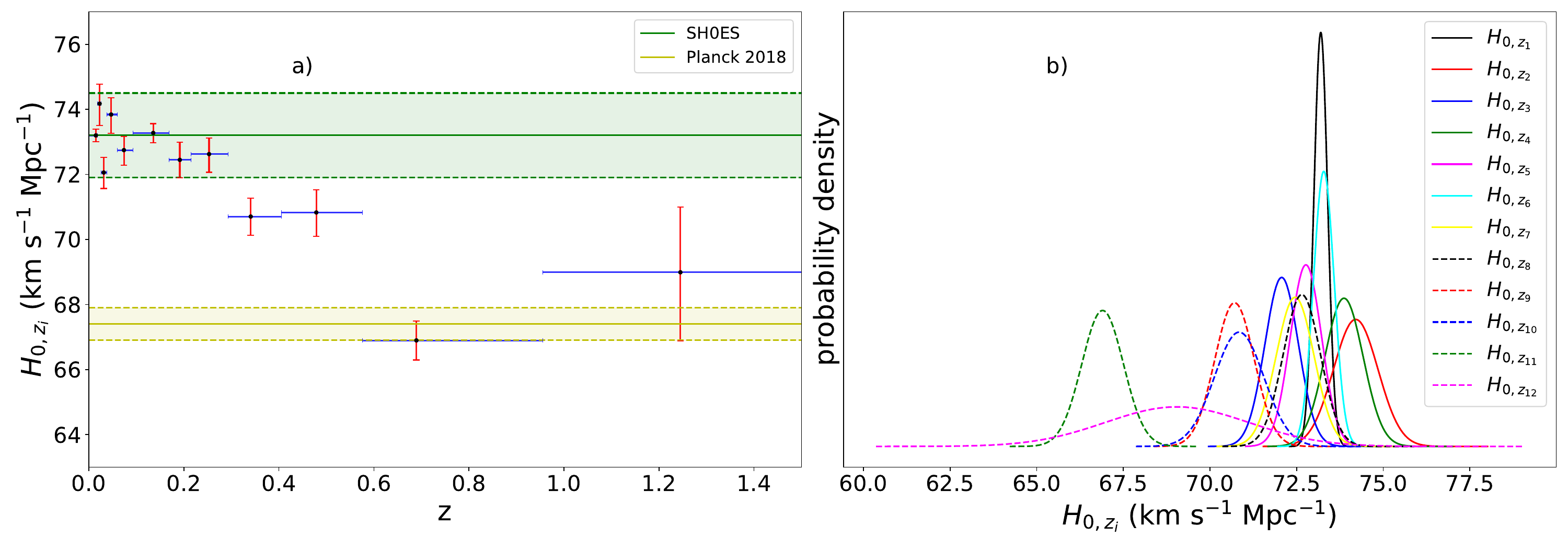}
    \caption{Fitting results of $H_{0,z}$ in eight redshift bins. Panel (a) shows the value of $H_{0,z}$ as a function of redshift. There is a clear decreasing trend with $4.9 \sigma$ significance between $z = 0$ to $z = 1.5$.  The green line gives $H_{0} = 73.04\pm 1.04$ km s$^{-1}$ Mpc$^{-1}$ from the local distance ladder and its 1$\sigma$ uncertainty \citep{Riess2022}. The yellow line is the value of $H_{0} = 67.4\pm 0.5$ km s$^{-1}$ Mpc$^{-1}$ from the CMB measurements and its 1$\sigma$ uncertainty \citep{PlanckCollaboration2020}. Panel (b) shows the probability density of $H_{0,z}$.}
    \label{F_bb}
\end{figure*}

In addition, we conduct a detailed study on different binning methods. We first group the sample based on their numbers to ensure that the statistical significance of each bin is nearly consistent (referred to as Model 1). The upper boundaries of these seven bins are $z_i = 0.024, 0.038, 0.14, 0.23, 0.48$ and $1.5$. The results shown in Table \ref{T_Model1} demonstrate that the error of the fitting results for each bin is nearly identical. Moreover, the decreasing trend still persists. We also try another binning method (referred to as Model2), the results are shown in Table \ref{T_Model2}. As shown in Figure \ref{Fbin7}, even with different binning methods, there is still evidence of the evolution of $H_{0,z}$. For these two binning methods, the decreasing trend obtained by using the calculation mentioned above are $7.4 \sigma$ and $3.8 \sigma$, respectively. Moreover, to test the robustness of our findings, we employ the Bayesian block methodology for data binning. This approach allows for a more nuanced and statistically sound analysis of the dataset. The binning boundaries and the corresponding data within each bin are given in table \ref{T_BB}. The results, as illustrated in Figure \ref{F_bb}, demonstrate a statistically significant decreasing trend with a confidence level of $4.9 \sigma$. Notably, all three distinct binning methods consistently reveal similar decreasing trends for $H_{0,z}$, supporting that the decline of $H_{0,z}$ is not an artifact of the binning method.

\begin{figure*}
    \centering
    \includegraphics[width=\textwidth,angle=0]{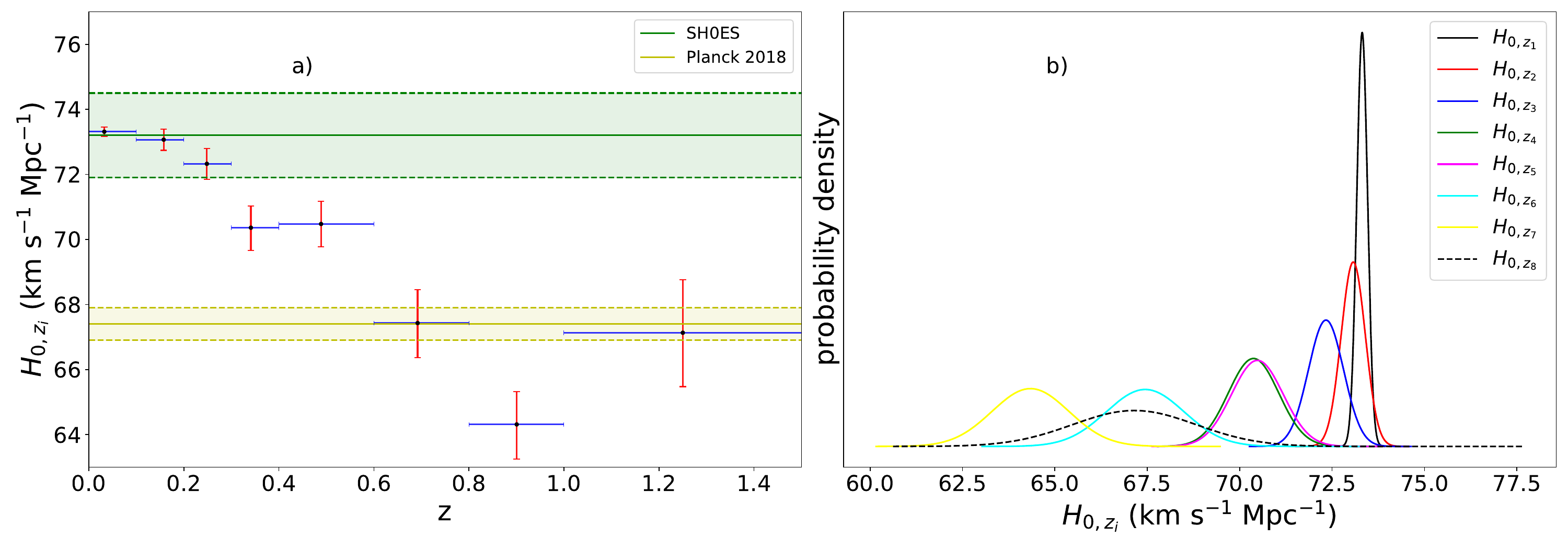}
    \caption{Fitting results of $H_{0,z}$ in eight redshift bins. Panel (a) shows the value of $H_{0,z}$ as a function of redshift. There is a clear decreasing trend with $7.2 \sigma$ significance between $z = 0$ to $z = 1.5$.  The green line gives $H_{0} = 73.04\pm 1.04$ km s$^{-1}$ Mpc$^{-1}$ from the local distance ladder and its 1$\sigma$ uncertainty \citep{Riess2022}. The yellow line is the value of $H_{0} = 67.4\pm 0.5$ km s$^{-1}$ Mpc$^{-1}$ from the CMB measurements and its 1$\sigma$ uncertainty \citep{PlanckCollaboration2020}. Panel (b) shows the probability density of $H_{0,z}$.}
    \label{F_noCC}
\end{figure*}

Considering that the Hubble parameter sample may not be a good sample, as differential ages suffer from massive systematics, we perform the fit without the Hubble parameter sample to avoid the possibility that the evolution of $H_{0,z}$ is caused by systematic effects in the dataset. The results are shown in Table \ref{T_noCC} and Figure \ref{F_noCC}. The decreasing trend is about $7.2 \sigma$, which is more significant than that of whole smaple. These two results are consistent within $1 \sigma$ confidence level, which rules out the systematics in the Hubble parameter sample as the cause of the evolution of $H_{0,z}$.

We also fit the flat $\Lambda$CDM model using the aforementioned sample, considering the same case with $\Omega_m$ fixed to 0.315. The result is $H_0 = 73.08 \pm 0.12$ km~\textrm{s$^{-1}$} \textrm{Mpc$^{-1}$}. Some standard information criteria are considered, such as the Akaike Information Criterion (AIC) \citep{Akaike1974}, the Bayesian Information Criterion (BIC) \citep{Schwarz1974} and the Deviance Information Criterion (DIC) \citep{DIC}. They are defined as
\begin{equation} \nonumber
    \text{AIC} = 2n - 2\ln(\mathcal{L}_\textrm{{max}}),
    \text{BIC} = n\ln(N) - 2\ln(\mathcal{L}_\textrm{{max}}),
\end{equation}
and
\begin{equation}
    \text{DIC} = 2n_{\textrm{eff}} - 2\ln(\mathcal{L}_\textrm{{max}}),
\end{equation}
where $\mathcal{L}_\textrm{{max}}$ is the maximum likelihood estimation, $n$ is the number of parameters, $N$ is the number of data and $n_\textrm{{eff}}$ represents the effective number of parameters. The comparison between the flat $\Lambda$CDM model and the $H_{0,z}$ model is presented in Table \ref{AICBIC}. The $\Delta$AIC value is calculated by subtracting the AIC value of the flat $\Lambda$CDM model from the AIC value of $H_{0,z}$ model, and the same approach is applied to compute $\Delta$BIC and $\Delta$DIC. 
Negative values indicate that the model being evaluated fits the data set better than the $\Lambda$CDM model, while positive values indicate a worse fit. The absolute values of the results exceed 10, indicating a very strong evidence against the flat $\Lambda$CDM model.

\section{Conclusions and Discussion}\label{Conclusion and Discussion}
Our results demonstrate an evident decreasing $H_0$ trend with a $6.4 \sigma$ significance. The decreasing trend was also discovered in the $H_0$ measured by gravitationally lensed objects with a lower significance ($\sim 2\sigma$) \citep{Wong2020,Kelly2023}. By including the new BAO data from DESI, the results are tighter compared to previous ones \citep{2023A&A...674A..45J}. The decreasing trend of the Hubble constant is more pronounced at low redshifts but tends to flatten at high redshifts. 

The combined data from different probes help to provide strong constraints on $H_0$. Considering the current scarcity of high-redshift data points, it is difficult to choose redshift bins at high redshifts. Here, we discard high-redshift data to focus on $z < 1.5$. There is no doubt that the high-redshift dataset provides strong constraints on $H_0$ \citep{Wu2022,Jia2022,2022ApJ...924...97W,2023ApJ...951...63D,2023MNRAS.525.3104B,2024arXiv241003994G}. The Pantheon+ SNe Ia sample is based on the Cepheid calibration \citep{2022ApJ...938..113S}.
In addition to the values obtained from the local distance ladder, the values calibrated via TRGB present $H_0 = 69.8 \pm 1.7$ km~\textrm{s$^{-1}$} \textrm{Mpc$^{-1}$} \citep{Freedman2021}. 
\cite{Cao2023} used a joint sample included Pantheon+, BAO, Hubble parameter data, quasar data, HII starburst galaxy, and gamma-ray burst to constrain cosmological parameters as $H_0 = 69.5 \pm 2.4$ km~\textrm{s$^{-1}$} \textrm{Mpc$^{-1}$}. The effect of the absolute magnitude prior on the estimation of $H_0$ is discussed in \cite{2024ApJ...964L...4C}. In our previous work \citep{2023A&A...674A..45J}, the decreasing trend of $H_{0,z}$ still persists when only SNe Ia sample is used. It demonstrates that the decreasing trend is not a statistical effect caused by the combination of different samples, but a genuinely existing phenomenon.

The impact of different parameters is considered. We first study the impact of $\Omega_m$ on the evolution of $H_{0, z}$. Since the quality of the current sample is insufficient to simultaneously constrain these parameters, we have to adopt the approach of fixing $\Omega_m$ as different values to study the variation of $H_{0, z}$. The results show that although the values of $H_{0, z}$ are influenced by the choice of $\Omega_m$ the evolutionary trend remains unchanged. In addition, we explore the impact of different binning methods on $H_{0, z}$. To determine whether the trend is real or a result of statistical effects caused by binning, we employ different binning methods for further analysis. Both of these binning methods demonstrate the decreasing trend. The significance is $7.4 \sigma$, $3.8 \sigma$, and $4.9 \sigma$ respectively, indicating that the trend is not caused by statistical effects of the binning method. What is more, we further analyze the sample used during the fitting process and discuss the potential bias in the calibration. 

The correlation among $H_{0,z_i}$ has been considered in this work. With the non-parametric method proposed by \cite{2005PhRvD..71b3506H} and \cite{Riess2007}, we estimate the Hubble constant at different redshift bins. The principal component analysis removes the correlation among different bins by diagonalizing the covariance matrix \citep{2005PhRvD..71b3506H,Riess2007}. It is noteworthy that some previous works do not consider the correlation among different bins, which may have caused discrepancies between their assumptions and methods. Nevertheless, their results also show a $H_0$ decreasing trend. To some extent, our non-parametric method is a good choice since it does not require adding prior conditions.

For the deviation from flat $\Lambda$CDM model, the reason may be due to systematic uncertainties in the data sample or the physical mechanism not captured by the current model. It is not sufficient to solve the deviation by invoking early-time new physics \citep{2023Univ....9..393V}. It is noteworthy that the diverse tracers of the BAO exhibit widely different linear bias parameters, which may evolve at different rates. This variation may bias the estimate of $H_0$ from BAO. Considering that the SNe Ia sample constitutes the majority of the data used, their systematic uncertainties, which may introduce bias into the results, should be discussed. The effect of different components of systematic uncertainties on the estimation of $H_0$ is extensively discussed in \cite{Brout2022}. The entire contribution of the uncertainty is below $0.7~\textrm{km}~\textrm{s}^{-1} \textrm{Mpc}^{-1}$, which is insufficient to explain the currently derived decreasing trend in $H_0$. However, the evolution of the light curve stretch may affect the value of the Hubble constant \citep{Nicolas2021}. On the other hand, there are still debates regarding the calibration of SNe Ia, which could lead to different inferred values of $H_0$ \citep{2023JCAP...11..050F}. One of the calibration methods is the Cepheid period-luminosity relation. In the construction of the local distance ladder, Cepheid variables occupy the first rung. Great caution needed to be exercised on the uncertainties at local universe, as small deviations at low redshift can lead to significant tension at high redshift. Fortunately, the hypothesis of unrecognized crowding of Cepheid photometry as the cause of the ``Hubble tension" is rejected by the latest observation from James Webb Space Telescope \citep{2024ApJ...962L..17R}. However, the transition behavior in the parameters of the Cepheid period-luminosity relation will affect the value of the Hubble constant derived from SNe Ia, especially since Cepheids occupy the first position in the distance ladder \citep{2021PhRvD.104l3511P,2022Univ....8..502P}. If the systematic uncertainties are reasonably estimated, our results support that the Hubble constant evolves with redshift. Some new physics, such as dynamical dark energy \citep{Ratra1988,2017NatAs...1..627z,2022PhRvD.106e5014Y,Cao2023,2024JHEP...05..327Y}, local voids \citep{2022MNRAS.511.5742W,2024MNRAS.tmp.2636M}, or modified gravity models \citep{Capozziello2011}, will have their moment on the stage. The standard cosmological $\Lambda$CDM model, long regarded as the ultimate reference, may soon undergo revisions.

\section*{acknowledgments}
We thank two anonymous referees for helpful comments. This work was supported by the National Natural Science Foundation of China
(grant Nos. 12494575 and 12273009), and the China Manned Spaced Project (CMS-CSST-2021-A12).

\bibliography{ms}{}

\begin{thebibliography}{}
\expandafter\ifx\csname natexlab\endcsname\relax\def\natexlab#1{#1}\fi
\providecommand{\url}[1]{\href{#1}{#1}}
\providecommand{\dodoi}[1]{doi:~\href{http://doi.org/#1}{\nolinkurl{#1}}}
\providecommand{\doeprint}[1]{\href{http://ascl.net/#1}{\nolinkurl{http://ascl.net/#1}}}
\providecommand{\doarXiv}[1]{\href{https://arxiv.org/abs/#1}{\nolinkurl{https://arxiv.org/abs/#1}}}

\bibitem[{Akaike(1974)}]{Akaike1974}
Akaike, H. 1974, IEEE Transactions on Automatic Control, 19, 716,
  \dodoi{10.1109/TAC.1974.1100705}

\bibitem[{Alam {et~al.}(2021)Alam, , {et~al.}}]{2021PhRvD.103h3533A}
Alam, S., , {et~al.} 2021, \prd, 103, 083533,
  \dodoi{10.1103/PhysRevD.103.083533}

\bibitem[{{Bargiacchi} {et~al.}(2023){Bargiacchi}, {Dainotti}, \&
  {Capozziello}}]{2023MNRAS.525.3104B}
{Bargiacchi}, G., {Dainotti}, M.~G., \& {Capozziello}, S. 2023, \mnras, 525,
  3104, \dodoi{10.1093/mnras/stad2326}

\bibitem[{{Bousis} \& {Perivolaropoulos}(2024)}]{Bousis2024}
{Bousis}, D., \& {Perivolaropoulos}, L. 2024, arXiv e-prints, arXiv:2405.07039,
  \dodoi{10.48550/arXiv.2405.07039}

\bibitem[{{Brout} {et~al.}(2022){Brout}, {Scolnic}, {Popovic}, {Riess}, {Carr},
  {Zuntz}, {Kessler}, {Davis}, {Hinton}, {Jones}, {Kenworthy}, {Peterson},
  {Said}, {Taylor}, {Ali}, {Armstrong}, {Charvu}, {Dwomoh}, {Meldorf},
  {Palmese}, {Qu}, {Rose}, {Sanchez}, {Stubbs}, {Vincenzi}, {Wood}, {Brown},
  {Chen}, {Chambers}, {Coulter}, {Dai}, {Dimitriadis}, {Filippenko}, {Foley},
  {Jha}, {Kelsey}, {Kirshner}, {M{\"o}ller}, {Muir}, {Nadathur}, {Pan}, {Rest},
  {Rojas-Bravo}, {Sako}, {Siebert}, {Smith}, {Stahl}, \& {Wiseman}}]{Brout2022}
{Brout}, D., {Scolnic}, D., {Popovic}, B., {et~al.} 2022, \apj, 938, 110,
  \dodoi{10.3847/1538-4357/ac8e04}

\bibitem[{{Calderon} {et~al.}(2024){Calderon}, {Lodha}, {Shafieloo}, {Linder},
  {Sohn}, {de Mattia}, {Cervantes-Cota}, {Crittenden}, {Davis}, {Ishak}, {Kim},
  {Matthewson}, {Niz}, {Park}, {Aguilar}, {Ahlen}, {Allen}, {Brooks},
  {Claybaugh}, {de la Macorra}, {Dey}, {Dey}, {Doel}, {Forero-Romero},
  {Gazta{\~n}aga}, {Gontcho}, {Honscheid}, {Howlett}, {Juneau}, {Kremin},
  {Landriau}, {Le Guillou}, {Levi}, {Manera}, {Miquel}, {Moustakas}, {Newman},
  {Palanque-Delabrouille}, {Percival}, {Poppett}, {Prada}, {Rezaie}, {Rossi},
  {Ruhlmann-Kleider}, {Sanchez}, {Schlegel}, {Schubnell}, {Seo}, {Sprayberry},
  {Tarl{\'e}}, {Taylor}, {Vargas-Maga{\~n}a}, {Weaver}, {Zarrouk}, \&
  {Zou}}]{2024arXiv240504216C}
{Calderon}, R., {Lodha}, K., {Shafieloo}, A., {et~al.} 2024, arXiv e-prints,
  arXiv:2405.04216, \dodoi{10.48550/arXiv.2405.04216}

\bibitem[{{Cao} \& {Ratra}(2022)}]{2022MNRAS.513.5686C}
{Cao}, S., \& {Ratra}, B. 2022, \mnras, 513, 5686,
  \dodoi{10.1093/mnras/stac1184}

\bibitem[{{Cao} \& {Ratra}(2023)}]{Cao2023}
---. 2023, \prd, 107, 103521, \dodoi{10.1103/PhysRevD.107.103521}

\bibitem[{{Capozziello} \& {de Laurentis}(2011)}]{Capozziello2011}
{Capozziello}, S., \& {de Laurentis}, M. 2011, \physrep, 509, 167,
  \dodoi{10.1016/j.physrep.2011.09.003}

\bibitem[{{Carloni} {et~al.}(2024){Carloni}, {Luongo}, \&
  {Muccino}}]{2024arXiv240412068C}
{Carloni}, Y., {Luongo}, O., \& {Muccino}, M. 2024, arXiv e-prints,
  arXiv:2404.12068, \dodoi{10.48550/arXiv.2404.12068}

\bibitem[{{Chen} {et~al.}(2024){Chen}, {Kumar}, {Ratra}, \&
  {Xu}}]{2024ApJ...964L...4C}
{Chen}, Y., {Kumar}, S., {Ratra}, B., \& {Xu}, T. 2024, \apjl, 964, L4,
  \dodoi{10.3847/2041-8213/ad2e97}

\bibitem[{{Colg{\'a}in} {et~al.}(2024){Colg{\'a}in}, {Dainotti}, {Capozziello},
  {Pourojaghi}, {Sheikh-Jabbari}, \& {Stojkovic}}]{2024arXiv240408633C}
{Colg{\'a}in}, E.~{\'O}., {Dainotti}, M.~G., {Capozziello}, S., {et~al.} 2024,
  arXiv e-prints, arXiv:2404.08633, \dodoi{10.48550/arXiv.2404.08633}

\bibitem[{{Dainotti} {et~al.}(2023){Dainotti}, {Bargiacchi}, {Bogdan},
  {Lenart}, {Iwasaki}, {Capozziello}, {Zhang}, \&
  {Fraija}}]{2023ApJ...951...63D}
{Dainotti}, M.~G., {Bargiacchi}, G., {Bogdan}, M., {et~al.} 2023, \apj, 951,
  63, \dodoi{10.3847/1538-4357/acd63f}

\bibitem[{{Dainotti} {et~al.}(2021){Dainotti}, {De Simone}, {Schiavone},
  {Montani}, {Rinaldi}, \& {Lambiase}}]{2021ApJ...912..150D}
{Dainotti}, M.~G., {De Simone}, B., {Schiavone}, T., {et~al.} 2021, \apj, 912,
  150, \dodoi{10.3847/1538-4357/abeb73}

\bibitem[{{Dainotti} {et~al.}(2022){Dainotti}, {De Simone}, {Schiavone},
  {Montani}, {Rinaldi}, {Lambiase}, {Bogdan}, \& {Ugale}}]{2022Galax..10...24D}
{Dainotti}, M.~G., {De Simone}, B.~D., {Schiavone}, T., {et~al.} 2022,
  Galaxies, 10, 24, \dodoi{10.3390/galaxies1001002410.48550/arXiv.2201.09848}

\bibitem[{{DESI Collaboration} {et~al.}(2024){DESI Collaboration}, {Adame},
  {Aguilar}, {Ahlen}, {Alam}, {Alexander}, {Alvarez}, {Alves}, {Anand},
  {Andrade}, {Armengaud}, {Avila}, {Aviles}, {Awan}, {Bahr-Kalus}, {Bailey},
  {Baltay}, {Bault}, {Behera}, {BenZvi}, {Bera}, {Beutler}, {Bianchi}, {Blake},
  {Blum}, {Brieden}, {Brodzeller}, {Brooks}, {Buckley-Geer}, {Burtin},
  {Calderon}, {Canning}, {Carnero Rosell}, {Cereskaite}, {Cervantes-Cota},
  {Chabanier}, {Chaussidon}, {Chaves-Montero}, {Chen}, {Chen}, {Claybaugh},
  {Cole}, {Cuceu}, {Davis}, {Dawson}, {de la Macorra}, {de Mattia}, {Deiosso},
  {Dey}, {Dey}, {Ding}, {Doel}, {Edelstein}, {Eftekharzadeh}, {Eisenstein},
  {Elliott}, {Fagrelius}, {Fanning}, {Ferraro}, {Ereza}, {Findlay}, {Flaugher},
  {Font-Ribera}, {Forero-S{\'a}nchez}, {Forero-Romero}, {Frenk},
  {Garcia-Quintero}, {Gazta{\~n}aga}, {Gil-Mar{\'\i}n}, {Gontcho},
  {Gonzalez-Morales}, {Gonzalez-Perez}, {Gordon}, {Green}, {Gruen}, {Gsponer},
  {Gutierrez}, {Guy}, {Hadzhiyska}, {Hahn}, {Hanif}, {Herrera-Alcantar},
  {Honscheid}, {Howlett}, {Huterer}, {Ir{\v{s}}i{\v{c}}}, {Ishak}, {Juneau},
  {Kara{\c{c}}ayl{\i}}, {Kehoe}, {Kent}, {Kirkby}, {Kremin}, {Krolewski},
  {Lai}, {Lan}, {Landriau}, {Lang}, {Lasker}, {Le Goff}, {Le Guillou},
  {Leauthaud}, {Levi}, {Li}, {Linder}, {Lodha}, {Magneville}, {Manera},
  {Margala}, {Martini}, {Maus}, {McDonald}, {Medina-Varela}, {Meisner},
  {Mena-Fern{\'a}ndez}, {Miquel}, {Moon}, {Moore}, {Moustakas}, {Mudur},
  {Mueller}, {Mu{\~n}oz-Guti{\'e}rrez}, {Myers}, {Nadathur}, {Napolitano},
  {Neveux}, {Newman}, {Nguyen}, {Nie}, {Niz}, {Noriega}, {Padmanabhan},
  {Paillas}, {Palanque-Delabrouille}, {Pan}, {Penmetsa}, {Percival}, {Pieri},
  {Pinon}, {Poppett}, {Porredon}, {Prada}, {P{\'e}rez-Fern{\'a}ndez},
  {P{\'e}rez-R{\`a}fols}, {Rabinowitz}, {Raichoor}, {Ram{\'\i}rez-P{\'e}rez},
  {Ramirez-Solano}, {Ravoux}, {Rashkovetskyi}, {Rezaie}, {Rich}, {Rocher},
  {Rockosi}, {Roe}, {Rosado-Marin}, {Ross}, {Rossi}, {Ruggeri},
  {Ruhlmann-Kleider}, {Samushia}, {Sanchez}, {Saulder}, {Schlafly}, {Schlegel},
  {Schubnell}, {Seo}, {Shafieloo}, {Sharples}, {Silber}, {Slosar}, {Smith},
  {Sprayberry}, {Tan}, {Tarl{\'e}}, {Taylor}, {Trusov}, {Ure{\~n}a-L{\'o}pez},
  {Vaisakh}, {Valcin}, {Valdes}, {Vargas-Maga{\~n}a}, {Verde}, {Walther},
  {Wang}, {Wang}, {Weaver}, {Weaverdyck}, {Wechsler}, {Weinberg}, {White},
  {Yu}, {Yu}, {Yuan}, {Y{\`e}che}, {Zaborowski}, {Zarrouk}, {Zhang}, {Zhao},
  {Zhao}, {Zhou}, {Zhuang}, \& {Zou}}]{2024arXiv240403002D}
{DESI Collaboration}, {Adame}, A.~G., {Aguilar}, J., {et~al.} 2024, arXiv
  e-prints, arXiv:2404.03002, \dodoi{10.48550/arXiv.2404.03002}

\bibitem[{{Farooq} {et~al.}(2013){Farooq}, {Crandall}, \&
  {Ratra}}]{2013PhLB..726...72F}
{Farooq}, O., {Crandall}, S., \& {Ratra}, B. 2013, Physics Letters B, 726, 72,
  \dodoi{10.1016/j.physletb.2013.08.078}

\bibitem[{{Farooq} {et~al.}(2017){Farooq}, {Ranjeet Madiyar}, {Crandall}, \&
  {Ratra}}]{2017ApJ...835...26F}
{Farooq}, O., {Ranjeet Madiyar}, F., {Crandall}, S., \& {Ratra}, B. 2017, \apj,
  835, 26, \dodoi{10.3847/1538-4357/835/1/26}

\bibitem[{{Foreman-Mackey} {et~al.}(2013){Foreman-Mackey}, {Hogg}, {Lang}, \&
  {Goodman}}]{Foreman-Mackey2013}
{Foreman-Mackey}, D., {Hogg}, D.~W., {Lang}, D., \& {Goodman}, J. 2013, \pasp,
  125, 306, \dodoi{10.1086/670067}

\bibitem[{{Freedman}(2021)}]{Freedman2021}
{Freedman}, W.~L. 2021, \apj, 919, 16, \dodoi{10.3847/1538-4357/ac0e95}

\bibitem[{{Freedman} \& {Madore}(2023)}]{2023JCAP...11..050F}
{Freedman}, W.~L., \& {Madore}, B.~F. 2023, \jcap, 2023, 050,
  \dodoi{10.1088/1475-7516/2023/11/050}

\bibitem[{{Gao} {et~al.}(2024){Gao}, {Wu}, {Hu}, {Yi}, {Zhou}, \&
  {Wang}}]{2024arXiv241003994G}
{Gao}, D.~H., {Wu}, Q., {Hu}, J.~P., {et~al.} 2024, arXiv e-prints,
  arXiv:2410.03994, \dodoi{10.48550/arXiv.2410.03994}

\bibitem[{{Hu} {et~al.}(2024{\natexlab{a}}){Hu}, {Jia}, {Hu}, \&
  {Wang}}]{2024ApJ...975L..36H}
{Hu}, J.~P., {Jia}, X.~D., {Hu}, J., \& {Wang}, F.~Y. 2024{\natexlab{a}},
  \apjl, 975, L36, \dodoi{10.3847/2041-8213/ad85cf}

\bibitem[{{Hu} \& {Wang}(2022)}]{Hu2022}
{Hu}, J.~P., \& {Wang}, F.~Y. 2022, \mnras, 517, 576,
  \dodoi{10.1093/mnras/stac2728}

\bibitem[{Hu \& Wang(2023)}]{universe9020094}
Hu, J.-P., \& Wang, F.-Y. 2023, Universe, 9, \dodoi{10.3390/universe9020094}

\bibitem[{{Hu} {et~al.}(2024{\natexlab{b}}){Hu}, {Wang}, {Hu}, \&
  {Wang}}]{2024A&A...681A..88H}
{Hu}, J.~P., {Wang}, Y.~Y., {Hu}, J., \& {Wang}, F.~Y. 2024{\natexlab{b}},
  \aap, 681, A88, \dodoi{10.1051/0004-6361/202347121}

\bibitem[{{Huterer} \& {Cooray}(2005)}]{2005PhRvD..71b3506H}
{Huterer}, D., \& {Cooray}, A. 2005, \prd, 71, 023506,
  \dodoi{10.1103/PhysRevD.71.023506}

\bibitem[{{Jia} {et~al.}(2023){Jia}, {Hu}, \& {Wang}}]{2023A&A...674A..45J}
{Jia}, X.~D., {Hu}, J.~P., \& {Wang}, F.~Y. 2023, \aap, 674, A45,
  \dodoi{10.1051/0004-6361/202346356}

\bibitem[{{Jia} {et~al.}(2022){Jia}, {Hu}, {Yang}, {Zhang}, \&
  {Wang}}]{Jia2022}
{Jia}, X.~D., {Hu}, J.~P., {Yang}, J., {Zhang}, B.~B., \& {Wang}, F.~Y. 2022,
  \mnras, 516, 2575, \dodoi{10.1093/mnras/stac2356}

\bibitem[{{Jimenez} \& {Loeb}(2002)}]{2002ApJ...573...37J}
{Jimenez}, R., \& {Loeb}, A. 2002, \apj, 573, 37, \dodoi{10.1086/340549}

\bibitem[{{Kazantzidis} \& {Perivolaropoulos}(2020)}]{Kazantzidis2020}
{Kazantzidis}, L., \& {Perivolaropoulos}, L. 2020, \prd, 102, 023520,
  \dodoi{10.1103/PhysRevD.102.023520}

\bibitem[{{Kelly} {et~al.}(2023){Kelly}, {Rodney}, {Treu}, {Oguri}, {Chen},
  {Zitrin}, {Birrer}, {Bonvin}, {Dessart}, {Diego}, {Filippenko}, {Foley},
  {Gilman}, {Hjorth}, {Jauzac}, {Mandel}, {Millon}, {Pierel}, {Sharon},
  {Thorp}, {Williams}, {Broadhurst}, {Dressler}, {Graur}, {Jha}, {McCully},
  {Postman}, {Schmidt}, {Tucker}, \& {von der Linden}}]{Kelly2023}
{Kelly}, P.~L., {Rodney}, S., {Treu}, T., {et~al.} 2023, Science, 380, abh1322,
  \dodoi{10.1126/science.abh1322}

\bibitem[{{Krishnan} {et~al.}(2020){Krishnan}, {Colg{\'a}in}, {Ruchika},
  {Sheikh-Jabbari}, \& {Yang}}]{2020PhRvD.102j3525K}
{Krishnan}, C., {Colg{\'a}in}, E.~{\'O}., {Ruchika}, Sen, A.~A.,
  {Sheikh-Jabbari}, M.~M., \& {Yang}, T. 2020, \prd, 102, 103525,
  \dodoi{10.1103/PhysRevD.102.103525}

\bibitem[{{Malekjani} {et~al.}(2024){Malekjani}, {Mc Conville}, {{\'O}
  Colg{\'a}in}, {Pourojaghi}, \& {Sheikh-Jabbari}}]{Malekjani2024}
{Malekjani}, M., {Mc Conville}, R., {{\'O} Colg{\'a}in}, E., {Pourojaghi}, S.,
  \& {Sheikh-Jabbari}, M.~M. 2024, European Physical Journal C, 84, 317,
  \dodoi{10.1140/epjc/s10052-024-12667-z}

\bibitem[{{Mazurenko} {et~al.}(2024){Mazurenko}, {Banik}, \&
  {Kroupa}}]{2024MNRAS.tmp.2636M}
{Mazurenko}, S., {Banik}, I., \& {Kroupa}, P. 2024, \mnras,
  \dodoi{10.1093/mnras/stae2758}

\bibitem[{{Millon} {et~al.}(2020){Millon}, {Galan}, {Courbin}, {Treu}, {Suyu},
  {Ding}, {Birrer}, {Chen}, {Shajib}, {Sluse}, {Wong}, {Agnello}, {Auger},
  {Buckley-Geer}, {Chan}, {Collett}, {Fassnacht}, {Hilbert}, {Koopmans},
  {Motta}, {Mukherjee}, {Rusu}, {Sonnenfeld}, {Spiniello}, \& {Van de
  Vyvere}}]{Millon2020}
{Millon}, M., {Galan}, A., {Courbin}, F., {et~al.} 2020, \aap, 639, A101,
  \dodoi{10.1051/0004-6361/201937351}

\bibitem[{{Moresco}(2023)}]{Moresco2023}
{Moresco}, M. 2023, arXiv e-prints, arXiv:2307.09501,
  \dodoi{10.48550/arXiv.2307.09501}

\bibitem[{{Moresco} {et~al.}(2020){Moresco}, {Jimenez}, {Verde}, {Cimatti}, \&
  {Pozzetti}}]{Moresco2020}
{Moresco}, M., {Jimenez}, R., {Verde}, L., {Cimatti}, A., \& {Pozzetti}, L.
  2020, \apj, 898, 82, \dodoi{10.3847/1538-4357/ab9eb0}

\bibitem[{{Nicolas} {et~al.}(2021){Nicolas}, {Rigault}, {Copin}, {Graziani},
  {Aldering}, {Briday}, {Kim}, {Nordin}, {Perlmutter}, \&
  {Smith}}]{Nicolas2021}
{Nicolas}, N., {Rigault}, M., {Copin}, Y., {et~al.} 2021, \aap, 649, A74,
  \dodoi{10.1051/0004-6361/202038447}

\bibitem[{{{\'O} Colg{\'a}in} {et~al.}(2022){{\'O} Colg{\'a}in},
  {Sheikh-Jabbari}, {Solomon}, {Bargiacchi}, {Capozziello}, {Dainotti}, \&
  {Stojkovic}}]{2022PhRvD.106d1301O}
{{\'O} Colg{\'a}in}, E., {Sheikh-Jabbari}, M.~M., {Solomon}, R., {et~al.} 2022,
  \prd, 106, L041301, \dodoi{10.1103/PhysRevD.106.L041301}

\bibitem[{{{\'O} Colg{\'a}in} {et~al.}(2024){{\'O} Colg{\'a}in},
  {Sheikh-Jabbari}, {Solomon}, {Dainotti}, \& {Stojkovic}}]{OColgain2024}
{{\'O} Colg{\'a}in}, E., {Sheikh-Jabbari}, M.~M., {Solomon}, R., {Dainotti},
  M.~G., \& {Stojkovic}, D. 2024, Physics of the Dark Universe, 44, 101464,
  \dodoi{10.1016/j.dark.2024.101464}

\bibitem[{{Pang} {et~al.}(2024){Pang}, {Zhang}, \& {Huang}}]{Pang2024}
{Pang}, Y.-H., {Zhang}, X., \& {Huang}, Q.-G. 2024, arXiv e-prints,
  arXiv:2408.14787, \dodoi{10.48550/arXiv.2408.14787}

\bibitem[{{Park} {et~al.}(2024){Park}, {de Cruz Perez}, \& {Ratra}}]{Park2024}
{Park}, C.-G., {de Cruz Perez}, J., \& {Ratra}, B. 2024, arXiv e-prints,
  arXiv:2405.00502, \dodoi{10.48550/arXiv.2405.00502}

\bibitem[{{Perivolaropoulos} \& {Skara}(2021)}]{2021PhRvD.104l3511P}
{Perivolaropoulos}, L., \& {Skara}, F. 2021, \prd, 104, 123511,
  \dodoi{10.1103/PhysRevD.104.123511}

\bibitem[{{Perivolaropoulos} \&
  {Skara}(2022{\natexlab{a}})}]{Perivolaropoulos2022}
---. 2022{\natexlab{a}}, New Astronomy Reviews, 95, 101659,
  \dodoi{10.1016/j.newar.2022.101659}

\bibitem[{{Perivolaropoulos} \&
  {Skara}(2022{\natexlab{b}})}]{2022Univ....8..502P}
---. 2022{\natexlab{b}}, Universe, 8, 502, \dodoi{10.3390/universe8100502}

\bibitem[{{Perlmutter} {et~al.}(1999){Perlmutter}, {Aldering}, {Goldhaber},
  {Knop}, {Nugent}, {Castro}, {Deustua}, {Fabbro}, {Goobar}, {Groom}, {Hook},
  {Kim}, {Kim}, {Lee}, {Nunes}, {Pain}, {Pennypacker}, {Quimby}, {Lidman},
  {Ellis}, {Irwin}, {McMahon}, {Ruiz-Lapuente}, {Walton}, {Schaefer}, {Boyle},
  {Filippenko}, {Matheson}, {Fruchter}, {Panagia}, {Newberg}, {Couch}, \&
  {Project}}]{1999ApJ...517..565P}
{Perlmutter}, S., {Aldering}, G., {Goldhaber}, G., {et~al.} 1999, \apj, 517,
  565, \dodoi{10.1086/307221}

\bibitem[{{Planck Collaboration} {et~al.}(2020){Planck Collaboration},
  {Aghanim}, {Akrami}, {Ashdown}, {Aumont}, {Baccigalupi}, {Ballardini},
  {Banday}, {Barreiro}, {Bartolo}, {Basak}, {Battye}, {Benabed}, {Bernard},
  {Bersanelli}, {Bielewicz}, {Bock}, {Bond}, {Borrill}, {Bouchet}, {Boulanger},
  {Bucher}, {Burigana}, {Butler}, {Calabrese}, {Cardoso}, {Carron},
  {Challinor}, {Chiang}, {Chluba}, {Colombo}, {Combet}, {Contreras}, {Crill},
  {Cuttaia}, {de Bernardis}, {de Zotti}, {Delabrouille}, {Delouis}, {Di
  Valentino}, {Diego}, {Dor{\'e}}, {Douspis}, {Ducout}, {Dupac}, {Dusini},
  {Efstathiou}, {Elsner}, {En{\ss}lin}, {Eriksen}, {Fantaye}, {Farhang},
  {Fergusson}, {Fernandez-Cobos}, {Finelli}, {Forastieri}, {Frailis},
  {Fraisse}, {Franceschi}, {Frolov}, {Galeotta}, {Galli}, {Ganga},
  {G{\'e}nova-Santos}, {Gerbino}, {Ghosh}, {Gonz{\'a}lez-Nuevo}, {G{\'o}rski},
  {Gratton}, {Gruppuso}, {Gudmundsson}, {Hamann}, {Handley}, {Hansen},
  {Herranz}, {Hildebrandt}, {Hivon}, {Huang}, {Jaffe}, {Jones}, {Karakci},
  {Keih{\"a}nen}, {Keskitalo}, {Kiiveri}, {Kim}, {Kisner}, {Knox},
  {Krachmalnicoff}, {Kunz}, {Kurki-Suonio}, {Lagache}, {Lamarre}, {Lasenby},
  {Lattanzi}, {Lawrence}, {Le Jeune}, {Lemos}, {Lesgourgues}, {Levrier},
  {Lewis}, {Liguori}, {Lilje}, {Lilley}, {Lindholm}, {L{\'o}pez-Caniego},
  {Lubin}, {Ma}, {Mac{\'\i}as-P{\'e}rez}, {Maggio}, {Maino}, {Mandolesi},
  {Mangilli}, {Marcos-Caballero}, {Maris}, {Martin}, {Martinelli},
  {Mart{\'\i}nez-Gonz{\'a}lez}, {Matarrese}, {Mauri}, {McEwen}, {Meinhold},
  {Melchiorri}, {Mennella}, {Migliaccio}, {Millea}, {Mitra},
  {Miville-Desch{\^e}nes}, {Molinari}, {Montier}, {Morgante}, {Moss}, {Natoli},
  {N{\o}rgaard-Nielsen}, {Pagano}, {Paoletti}, {Partridge}, {Patanchon},
  {Peiris}, {Perrotta}, {Pettorino}, {Piacentini}, {Polastri}, {Polenta},
  {Puget}, {Rachen}, {Reinecke}, {Remazeilles}, {Renzi}, {Rocha}, {Rosset},
  {Roudier}, {Rubi{\~n}o-Mart{\'\i}n}, {Ruiz-Granados}, {Salvati}, {Sandri},
  {Savelainen}, {Scott}, {Shellard}, {Sirignano}, {Sirri}, {Spencer},
  {Sunyaev}, {Suur-Uski}, {Tauber}, {Tavagnacco}, {Tenti}, {Toffolatti},
  {Tomasi}, {Trombetti}, {Valenziano}, {Valiviita}, {Van Tent}, {Vibert},
  {Vielva}, {Villa}, {Vittorio}, {Wandelt}, {Wehus}, {White}, {White},
  {Zacchei}, \& {Zonca}}]{PlanckCollaboration2020}
{Planck Collaboration}, {Aghanim}, N., {Akrami}, Y., {et~al.} 2020, \aap, 641,
  A6, \dodoi{10.1051/0004-6361/201833910}

\bibitem[{{Podariu} {et~al.}(2001){Podariu}, {Souradeep}, {Gott}, {Ratra}, \&
  {Vogeley}}]{2001ApJ...559....9P}
{Podariu}, S., {Souradeep}, T., {Gott}, J.~Richard, I., {Ratra}, B., \&
  {Vogeley}, M.~S. 2001, \apj, 559, 9, \dodoi{10.1086/322409}

\bibitem[{{Ratra} \& {Peebles}(1988)}]{Ratra1988}
{Ratra}, B., \& {Peebles}, P.~J.~E. 1988, \prd, 37, 3406,
  \dodoi{10.1103/PhysRevD.37.3406}

\bibitem[{{Riess}(2020)}]{Riess2020}
{Riess}, A.~G. 2020, Nature Reviews Physics, 2, 10,
  \dodoi{10.1038/s42254-019-0137-0}

\bibitem[{{Riess} {et~al.}(1998){Riess}, {Filippenko}, {Challis},
  {Clocchiatti}, {Diercks}, {Garnavich}, {Gilliland}, {Hogan}, {Jha},
  {Kirshner}, {Leibundgut}, {Phillips}, {Reiss}, {Schmidt}, {Schommer},
  {Smith}, {Spyromilio}, {Stubbs}, {Suntzeff}, \&
  {Tonry}}]{1998AJ....116.1009R}
{Riess}, A.~G., {Filippenko}, A.~V., {Challis}, P., {et~al.} 1998, \aj, 116,
  1009, \dodoi{10.1086/300499}

\bibitem[{{Riess} {et~al.}(2007){Riess}, {Strolger}, {Casertano}, {Ferguson},
  {Mobasher}, {Gold}, {Challis}, {Filippenko}, {Jha}, {Li}, {Tonry}, {Foley},
  {Kirshner}, {Dickinson}, {MacDonald}, {Eisenstein}, {Livio}, {Younger}, {Xu},
  {Dahl{\'e}n}, \& {Stern}}]{Riess2007}
{Riess}, A.~G., {Strolger}, L.-G., {Casertano}, S., {et~al.} 2007, \apj, 659,
  98, \dodoi{10.1086/510378}

\bibitem[{{Riess} {et~al.}(2022){Riess}, {Yuan}, {Macri}, {Scolnic}, {Brout},
  {Casertano}, {Jones}, {Murakami}, {Anand}, {Breuval}, {Brink}, {Filippenko},
  {Hoffmann}, {Jha}, {D'arcy Kenworthy}, {Mackenty}, {Stahl}, \&
  {Zheng}}]{Riess2022}
{Riess}, A.~G., {Yuan}, W., {Macri}, L.~M., {et~al.} 2022, \apjl, 934, L7,
  \dodoi{10.3847/2041-8213/ac5c5b}

\bibitem[{{Riess} {et~al.}(2024){Riess}, {Anand}, {Yuan}, {Casertano},
  {Dolphin}, {Macri}, {Breuval}, {Scolnic}, {Perrin}, \&
  {Anderson}}]{2024ApJ...962L..17R}
{Riess}, A.~G., {Anand}, G.~S., {Yuan}, W., {et~al.} 2024, \apjl, 962, L17,
  \dodoi{10.3847/2041-8213/ad1ddd}

\bibitem[{Schwarz(1978)}]{Schwarz1974}
Schwarz, G. 1978, The Annals of Statistics, 6, 461 ,
  \dodoi{10.1214/aos/1176344136}

\bibitem[{{Scolnic} {et~al.}(2022){Scolnic}, {Brout}, {Carr}, {Riess}, {Davis},
  {Dwomoh}, {Jones}, {Ali}, {Charvu}, {Chen}, {Peterson}, {Popovic}, {Rose},
  {Wood}, {Brown}, {Chambers}, {Coulter}, {Dettman}, {Dimitriadis},
  {Filippenko}, {Foley}, {Jha}, {Kilpatrick}, {Kirshner}, {Pan}, {Rest},
  {Rojas-Bravo}, {Siebert}, {Stahl}, \& {Zheng}}]{2022ApJ...938..113S}
{Scolnic}, D., {Brout}, D., {Carr}, A., {et~al.} 2022, \apj, 938, 113,
  \dodoi{10.3847/1538-4357/ac8b7a}

\bibitem[{Spiegelhalter {et~al.}(2002)Spiegelhalter, Best, Carlin, \& Van
  Der~Linde}]{DIC}
Spiegelhalter, D.~J., Best, N.~G., Carlin, B.~P., \& Van Der~Linde, A. 2002,
  Journal of the Royal Statistical Society Series B: Statistical Methodology,
  64, 583, \dodoi{10.1111/1467-9868.00353}

\bibitem[{{Vagnozzi}(2023)}]{2023Univ....9..393V}
{Vagnozzi}, S. 2023, Universe, 9, 393, \dodoi{10.3390/universe9090393}

\bibitem[{{Verde} {et~al.}(2019){Verde}, {Treu}, \& {Riess}}]{Verde2019}
{Verde}, L., {Treu}, T., \& {Riess}, A.~G. 2019, Nature Astronomy, 3, 891,
  \dodoi{10.1038/s41550-019-0902-0}

\bibitem[{{Wang}(2024)}]{2024arXiv240413833W}
{Wang}, D. 2024, arXiv e-prints, arXiv:2404.13833,
  \dodoi{10.48550/arXiv.2404.13833}

\bibitem[{{Wang} {et~al.}(2022){Wang}, {Hu}, {Zhang}, \&
  {Dai}}]{2022ApJ...924...97W}
{Wang}, F.~Y., {Hu}, J.~P., {Zhang}, G.~Q., \& {Dai}, Z.~G. 2022, \apj, 924,
  97, \dodoi{10.3847/1538-4357/ac3755}

\bibitem[{{Wang} {et~al.}(2011){Wang}, {Qi}, \& {Dai}}]{Wang2011}
{Wang}, F.-Y., {Qi}, S., \& {Dai}, Z.-G. 2011, \mnras, 415, 3423,
  \dodoi{10.1111/j.1365-2966.2011.18961.x}

\bibitem[{{Wang} {et~al.}(2024{\natexlab{a}}){Wang}, {Peng}, \&
  {Piao}}]{WangH2024}
{Wang}, H., {Peng}, Z.-Y., \& {Piao}, Y.-S. 2024{\natexlab{a}}, arXiv e-prints,
  arXiv:2406.03395, \dodoi{10.48550/arXiv.2406.03395}

\bibitem[{{Wang} {et~al.}(2024{\natexlab{b}}){Wang}, {Lin}, {Ding}, \&
  {Hu}}]{2024arXiv240502168W}
{Wang}, Z., {Lin}, S., {Ding}, Z., \& {Hu}, B. 2024{\natexlab{b}}, arXiv
  e-prints, arXiv:2405.02168, \dodoi{10.48550/arXiv.2405.02168}

\bibitem[{{Wong} {et~al.}(2022){Wong}, {Shanks}, {Metcalfe}, \&
  {Whitbourn}}]{2022MNRAS.511.5742W}
{Wong}, J. H.~W., {Shanks}, T., {Metcalfe}, N., \& {Whitbourn}, J.~R. 2022,
  \mnras, 511, 5742, \dodoi{10.1093/mnras/stac396}

\bibitem[{{Wong} {et~al.}(2020)}]{Wong2020}
{Wong}, K.~C., {et~al.} 2020, \mnras, 498, 1420, \dodoi{10.1093/mnras/stz3094}

\bibitem[{{Wu} {et~al.}(2022){Wu}, {Zhang}, \& {Wang}}]{Wu2022}
{Wu}, Q., {Zhang}, G.-Q., \& {Wang}, F.-Y. 2022, \mnras, 515, L1,
  \dodoi{10.1093/mnrasl/slac022}

\bibitem[{{Yin}(2022)}]{2022PhRvD.106e5014Y}
{Yin}, W. 2022, \prd, 106, 055014, \dodoi{10.1103/PhysRevD.106.055014}

\bibitem[{{Yin}(2024)}]{2024JHEP...05..327Y}
---. 2024, Journal of High Energy Physics, 2024, 327,
  \dodoi{10.1007/JHEP05(2024)327}

\bibitem[{{Yu} {et~al.}(2018){Yu}, {Ratra}, \& {Wang}}]{2018ApJ...856....3Y}
{Yu}, H., {Ratra}, B., \& {Wang}, F.-Y. 2018, \apj, 856, 3,
  \dodoi{10.3847/1538-4357/aab0a2}

\bibitem[{{Zhao} {et~al.}(2017)}]{2017NatAs...1..627z}
{Zhao}, G.-B., {et~al.} 2017, Nature Astronomy, 1, 627,
  \dodoi{10.1038/s41550-017-0216-z}

\end{thebibliography}
\bibliographystyle{aasjournal}

\clearpage

\begin{table}\large
	\caption{Fitting results of $H_{0,z_i}$ (in units of $\textrm{km}~\textrm{s}^{-1} \textrm{Mpc}^{-1}$). \label{T_H8}}
	\centering
	\begin{tabular}{ccccc}
		\\
		\hline
		Redshift bin   & Number of SNe Ia & Number of H(z) & Number of BAO  &$H_{0,z_i}$   \\
		\hline
		$[0,0.10]$     & 630              &2               & 0              &$73.28^{+0.14}_{-0.14}$ \\
		$[0.10,0.20]$  & 207              &4               & 1              &$72.96^{+0.31}_{-0.31}$ \\
		$[0.20,0.30]$  & 259              &3               & 0              &$72.39^{+0.47}_{-0.46}$ \\
		$[0.30,0.40]$  & 186              &2               & 2              &$70.49^{+0.66}_{-0.66}$ \\
		$[0.40,0.60]$  & 179              &8               & 2              &$70.52^{+0.67}_{-0.63}$ \\
		$[0.60,0.80]$  & 99               &4               & 2              &$67.33^{+1.00}_{-0.99}$ \\
		$[0.80,1.00]$  & 5                &4               & 2              &$64.85^{+1.01}_{-1.00}$ \\
		$[1.00,1.50]$  & 18               &5               & 3              &$68.14^{+1.64}_{-1.60}$ \\
		\hline
	\end{tabular}
\end{table}

\begin{table}\large
	\caption{The results of $H_{0,z_i}$ (in units of $\textrm{km}~\textrm{s}^{-1} \textrm{Mpc}^{-1}$) for different $\Omega_m$ values. \label{T_Om}}
	\centering
	\begin{tabular}{ccccccc}
		\\
		\hline
                        & $\Omega_m = 0.26$       & $\Omega_m = 0.28$       & $\Omega_m = 0.30$       & $\Omega_m = 0.32$       & $\Omega_m = 0.34$ & $\Omega_m = 0.36$ \\
            \hline
            $H_{0,z_1}$ & $73.74^{+0.14}_{-0.14}$ & $73.56^{+0.14}_{-0.14}$ & $73.44^{+0.14}_{-0.14}$ & $73.21^{+0.14}_{-0.14}$ & $73.15^{+0.14}_{-0.15}$ & $72.88^{+0.14}_{-0.14}$\\
            $H_{0,z_2}$ & $74.11^{+0.33}_{-0.32}$ & $73.68^{+0.31}_{-0.31}$ & $73.30^{+0.32}_{-0.32}$ & $72.85^{+0.31}_{-0.29}$ & $72.64^{+0.33}_{-0.32}$ & $72.18^{+0.32}_{-0.29}$ \\
            $H_{0,z_3}$ & $73.89^{+0.48}_{-0.47}$ & $73.34^{+0.48}_{-0.47}$ & $72.73^{+0.49}_{-0.46}$ & $72.26^{+0.48}_{-0.47}$ & $71.80^{+0.48}_{-0.48}$ & $71.28^{+0.48}_{-0.45}$\\
            $H_{0,z_4}$ & $72.41^{+0.69}_{-0.68}$ & $71.64^{+0.69}_{-0.64}$ & $70.76^{+0.70}_{-0.64}$ & $70.28^{+0.69}_{-0.69}$ & $69.52^{+0.69}_{-0.68}$ & $68.88^{+0.66}_{-0.63}$\\
            $H_{0,z_5}$ & $73.39^{+0.69}_{-0.69}$ & $72.31^{+0.71}_{-0.69}$ & $71.06^{+0.68}_{-0.65}$ & $70.23^{+0.65}_{-0.65}$ & $69.19^{+0.67}_{-0.65}$ & $68.45^{+0.68}_{-0.63}$\\
            $H_{0,z_6}$ & $71.11^{+1.07}_{-1.00}$ & $69.79^{+1.01}_{-0.98}$ & $68.19^{+1.03}_{-0.97}$ & $67.28^{+0.95}_{-0.95}$ & $65.56^{+0.99}_{-0.96}$ & $64.70^{+0.96}_{-0.94}$\\
            $H_{0,z_7}$ & $69.01^{+1.04}_{-1.04}$ & $67.29^{+1.06}_{-0.99}$ & $65.44^{+1.07}_{-1.05}$ & $64.36^{+1.02}_{-0.96}$ & $62.43^{+1.03}_{-1.00}$ & $61.29^{+1.06}_{-1.00}$\\
            $H_{0,z_8}$ & $73.60^{+1.69}_{-1.63}$ & $71.60^{+1.73}_{-1.63}$ & $69.44^{+1.75}_{-1.67}$ & $67.78^{+1.68}_{-1.64}$ & $65.62^{+1.73}_{-1.64}$ & $64.43^{+1.69}_{-1.66}$\\
		\hline
	\end{tabular}
\end{table}

\begin{table}\large
	\caption{Fitting results of $H_{0,z_i}$ (in units of $\textrm{km}~\textrm{s}^{-1} \textrm{Mpc}^{-1}$) for Model 1. \label{T_Model1}}
	\centering
	\begin{tabular}{ccccc}
		\\
		\hline
		Redshift bin   & Number of SNe Ia & Number of H(z) & Number of BAO  &$H_{0,z_i}$   \\
		\hline
		$[0,0.024]$      & 244              &0               & 0              &$73.39^{+0.19}_{-0.19}$ \\
		$[0.024,0.038]$  & 224              &0               & 0              &$72.64^{+0.58}_{-0.53}$ \\
		$[0.038,0.140]$  & 226              &3               & 1              &$73.28^{+0.23}_{-0.23}$ \\
		$[0.140,0.230]$  & 221              &4               & 0              &$72.94^{+0.38}_{-0.37}$ \\
		$[0.230,0.320]$  & 232              &2               & 0              &$72.24^{+0.59}_{-0.54}$ \\
		$[0.320,0.480]$  & 224              &9               & 2              &$70.75^{+0.63}_{-0.58}$ \\
		$[0.480,1.500]$  & 212              &14              & 9              &$68.09^{+0.48}_{-0.47}$ \\
		\hline
	\end{tabular}
\end{table}

\begin{table}\large
	\caption{Fitting results of $H_{0,z_i}$ (in units of $\textrm{km}~\textrm{s}^{-1} \textrm{Mpc}^{-1}$) for Model 2. \label{T_Model2}}
	\centering
	\begin{tabular}{ccccc}
		\\
		\hline
		Redshift bin   & Number of SNe Ia & Number of H(z) & Number of BAO  &$H_{0,z_i}$   \\
		\hline
		$[0,0.2]$    & 837              &7               & 1              &$73.22^{+0.13}_{-0.13}$ \\
		$[0.2,0.4]$  & 445              &5               & 2              &$71.58^{+0.38}_{-0.37}$ \\
		$[0.4,0.6]$  & 179              &7               & 2              &$70.26^{+0.72}_{-0.68}$ \\
		$[0.6,0.8]$  & 99               &5               & 2              &$67.13^{+1.02}_{-1.00}$ \\
		$[0.8,1.0]$  & 5                &3               & 2              &$64.42^{+1.03}_{-1.05}$ \\
		$[1.0,1.3]$  & 9                &3               & 0              &$70.31^{+3.39}_{-3.31}$ \\
		$[1.3,1.5]$  & 9                &2               & 3              &$66.93^{+2.13}_{-2.12}$ \\
		\hline
	\end{tabular}
\end{table}

\begin{table}\large
	\caption{Fitting results of $H_{0,z_i}$ (in units of $\textrm{km}~\textrm{s}^{-1} \textrm{Mpc}^{-1}$) for Bayesian block methodology. \label{T_BB}}
	\centering
	\begin{tabular}{ccccc}
		\\
		\hline
		Redshift bin         & Number of SNe Ia & Number of H(z) & Number of BAO  &$H_{0,z_i}$   \\
		\hline
		$[0,0.0181]$        & 137              &0               & 0              &$73.22^{+0.13}_{-0.13}$ \\
		$[0.0181,0.0259]$  & 151              &0               & 0              &$71.58^{+0.38}_{-0.37}$ \\
		$[0.0259,0.0376]$  & 177              &0               & 0              &$70.26^{+0.72}_{-0.68}$ \\
		$[0.0376,0.0603]$  & 102              &0               & 0              &$67.13^{+1.02}_{-1.00}$ \\
		$[0.0603,0.0925]$  & 61               &2               & 0              &$64.42^{+1.03}_{-1.05}$ \\
		$[0.0925,0.1688]$  & 118              &1               & 1              &$70.31^{+3.39}_{-3.31}$ \\
		$[0.1688,0.2150]$  & 130              &4               & 0              &$66.93^{+2.13}_{-2.12}$ \\
        $[0.2150,0.2930]$  & 206              &2               & 0              &$66.93^{+2.13}_{-2.12}$ \\
        $[0.2930,0.4052]$  & 206              &4               & 2              &$66.93^{+2.13}_{-2.12}$ \\
        $[0.4052,0.5756]$  & 152              &5               & 2              &$66.93^{+2.13}_{-2.12}$ \\
        $[0.5756,0.9550]$  & 124              &9               & 4              &$66.93^{+2.13}_{-2.12}$ \\
        $[0.9550,1.5]$      & 19               &2               & 3              &$66.93^{+2.13}_{-2.12}$ \\
		\hline
	\end{tabular}
\end{table}

\begin{table}\large
	\caption{Fitting results of $H_{0,z_i}$ (in units of $\textrm{km}~\textrm{s}^{-1} \textrm{Mpc}^{-1}$) without the Hubble parameter sample \label{T_noCC}}
	\centering
	\begin{tabular}{cccc}
		\\
		\hline
		Redshift bin   & Number of SNe Ia & Number of BAO  &$H_{0,z_i}$   \\
		\hline
		$[0,0.10]$     & 630              & 0              &$73.31^{+0.14}_{-0.14}$ \\
		$[0.10,0.20]$  & 207              & 1              &$73.07^{+0.32}_{-0.31}$ \\
		$[0.20,0.30]$  & 259              & 0              &$72.32^{+0.48}_{-0.47}$ \\
		$[0.30,0.40]$  & 186              & 2              &$70.36^{+0.69}_{-0.67}$ \\
		$[0.40,0.60]$  & 179              & 2              &$70.48^{+0.70}_{-0.70}$ \\
		$[0.60,0.80]$  & 99               & 2              &$67.43^{+1.06}_{-1.03}$ \\
		$[0.80,1.00]$  & 5                & 2              &$64.31^{+1.07}_{-1.01}$ \\
		$[1.00,1.50]$  & 18               & 3              &$67.13^{+1.66}_{-1.64}$ \\
		\hline
	\end{tabular}
\end{table}

\begin{table}
	\caption{Model comparison \label{AICBIC}}
	\centering
	\begin{tabular}{ccccccc} 
		\\
		\hline
		Model & AIC & $\Delta$AIC & BIC & $\Delta$BIC  & DIC & $\Delta$DIC\\
		\hline
		$\Lambda$CDM  & 1616.61 & 0 & 1622.01 & 0 & 1616.72 & 0\\
		$H_{0,z}$ model & 1519.01 & -97.6 & 1562.16 & -59.85 & 1508.89 & -107.83\\
		\hline
	\end{tabular}~~~~~~~~
\end{table}

\end{document}